# Water dimer driven DNA base superstructure with mismatched hydrogen-bonding

Shuning Cai,[†,¶] Lauri Kurki,[†,¶] Chen Xu,[†,¶] Adam S. Foster,[*,†,‡] and Peter Liljeroth[*,†]

[†]*Department of Applied Physics, Aalto University, 00076 Aalto, Espoo, Finland*

[‡]*WPI Nano Life Science Institute (WPI-NanoLSI), Kanazawa University, Kakuma-machi, Kanazawa 920-1192, Japan*

[¶]*These authors contributed equally.*

E-mail: adam.foster@aalto.fi; peter.liljeroth@aalto.fi

## Abstract

The existence of water dimers in equilibrium water vapor at room temperature and their anomalous properties revealed by recent studies suggest the benchmark role of water dimer in both experiment and theory. However, there has been a limited observation of individual water dimers due to the challenge of water separation and generation at the single-molecule level. Here, we achieve real-space imaging of individual confined water dimers embedded inside self-assembled layer of a DNA base, adenine, on Ag(111). The hydration of the adenine layers by these water dimers causes a local surface chiral inversion in a way that the neighboring homochiral adenine molecules become heterochiral after hydration, resulting in a mismatched hydrogen-bond pattern between neighboring adenine molecules. Furthermore, the mutual influence between the adenine superstructure and these dynamic confined water dimers is corroborated by theoretical simulation and calculations. The observation of single confined water dimers offers an unprecedented approach to studying the fundamental forms of water clusters and their interaction with the local chemical environment.

The ubiquitous character of water as a solvent in nature means that in most cases, reactants and products are inevitably exposed to water and it plays a key role in the biochemical processes of living organisms. Intriguingly, many recent studies have expanded the role of water beyond a passive matrix to an active promoter,[1–3] in chemical reactions[4–6] and functional materials.[7–9] In particular, water dimers are not only a fundamental unit for studying the properties of water, but also show some exotic properties such as an anomalously low barrier for diffusion on a surface involving nuclear quantum effects (NQEs).[10–12] Besides, water dimers have been also proved to be a potential bifunctional catalyst,[13,14] and play a significant role in the initial stage of the ice nanocluster formation during the bilayer ice growth.[15–17] However, separating and generating individual water dimers stabilized at ambient temperature remains challenging.[18] Furthermore, many experimental techniques suffer from limited spatial resolution, yield ensemble-averaged results, or require complicated modeling to extract structural data. Thus, high-resolution structural data on the single-water-molecule level[19–22] would significantly improve our understanding of water dimers.

A viable route to directly study the properties of water dimers is to take advantage of the confinement possibilities offered by molecular assemblies. Supramolecular networks held together by non-covalent interactions have been considered as an ideal model to gain insights into micro-hydration,[23,24] due to their sensitivity to the external environment and ability to provide dynamic confinement. More generally, the extensive hydrogen bond (H-bond) is one of the most significant properties of water and recently the role of confined water in DNA bases[25–27] has aroused a broad interest in the investigation of DNA-related biological processes in vivo, creating novel opportunities in DNA nanotechnology. DNA bases record the genetic information of life through the H-bonded pairing mechanism with high efficiency and precision, which are widely applied in patterning diverse materials such as carbon nanotubes[28,29] and enriching the crystal structures of nanoparticles.[30,31] As one of the four nucleobases, adenine and its derivatives perform multiple functions in biochemistry. On the surface, adenine molecules form a H-bonded assisted 2D supramolecular network,[32] which can be considered as a low-dimensional crystal,[33] providing a unique platform to observe a micro-hydrated environment.

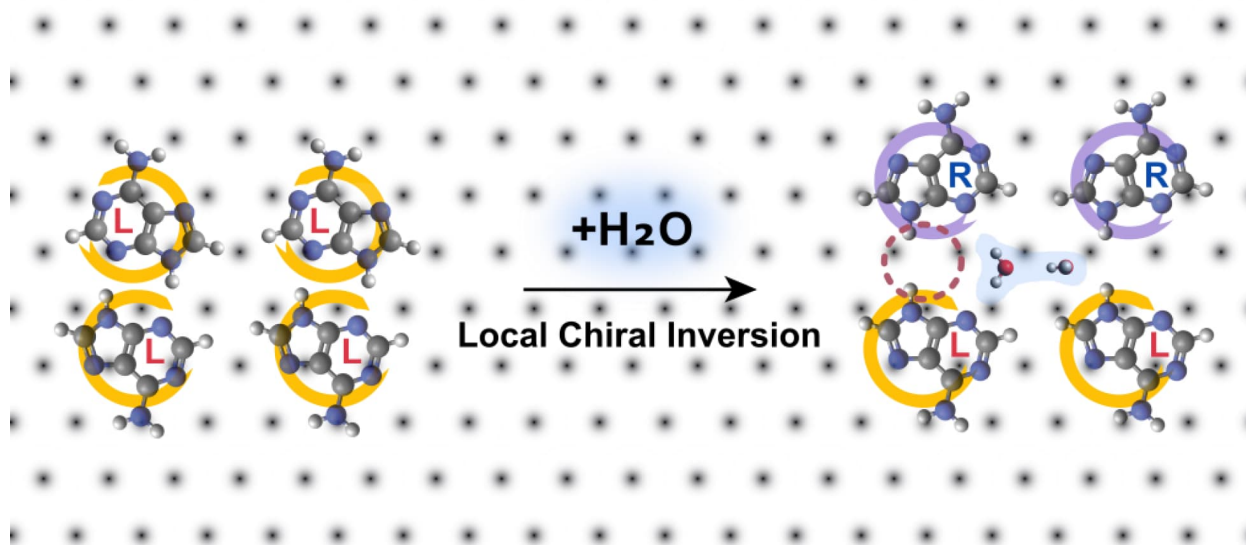

Scheme 1: Schematic of the adsorption site change, the local chiral inversion (purple and pink color represent left and right handedness, respectively) and the counter-intuitive H-bond 'mismatch' induced by the water dimers in a self-assembled adenine layer. The underlaying substrate (Ag(111)) atomic lattice is indicated by the black dots.

In this work, using low-temperature scanning tunneling microscopy (STM) and non-contact atomic force microscopy (nc-AFM) with a CO tip,[34,35] we achieve real-space imaging of water dimers within a dynamic 2D adenine layer on Ag(111). Instead of desorption or randomly distributed clustering, these dimers form spontaneously at room temperature (RT). As demonstrated in Scheme 1, the introduction of water dimers induces a local surface chiral inversion[36] in a way that the neighboring homochiral adenine pairs become heterochiral (left and right handedness marked by the purple and orange circular arrows, respectively), accompanied by the emergence of a counterintuitive mismatched H-bond structure (depicted by a red dashed circle). The necessity of the water dimer with a linear non-planar configuration in stabilizing the mismatched H-bonded superstructure is further confirmed by density functional theory (DFT) calculations. This direct observation provides rare evidence of the potential role of individual water dimers in self-assembly, paving the way for exploring the novel properties and the future applications of water dimers.

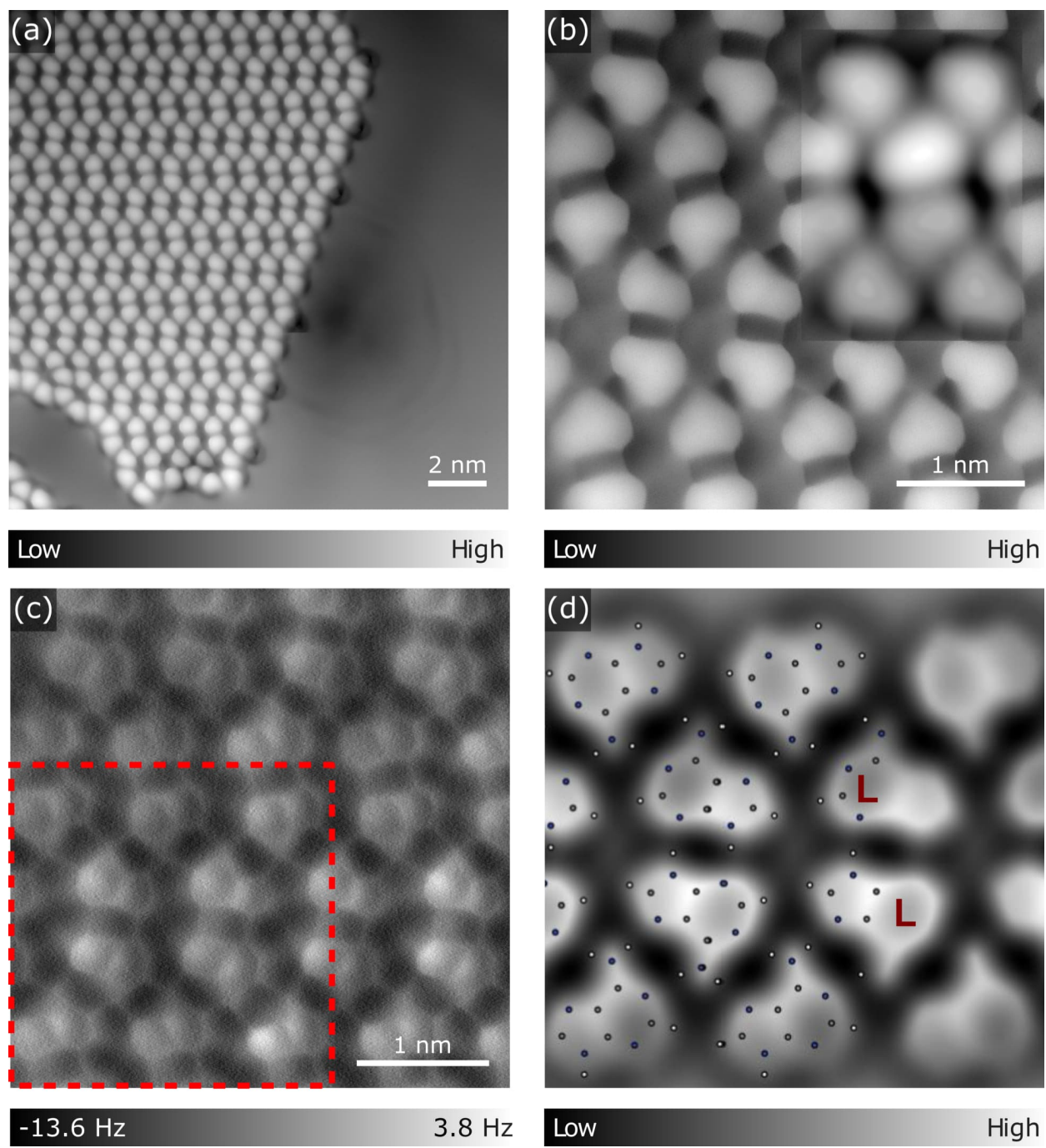

Figure 1: Self-assembly of adenine on Ag(111) probed by a CO-tip STM and AFM. (a) STM topography overview of an adenine island. Setpoint: 100 mV, 100 pA. (b,c) Zoomed-in STM topography of the island with a simulated STM image as an inset (b) and the corresponding constant-height AFM image (c). (d) Simulated AFM image with the overlaid molecular structure corresponding to the area marked in (c).

A H-bonded supramolecular network of adenine molecules on Ag(111) (Figure 1a) is formed at RT under ultra-high vacuum (details in the Supporting Information (SI)). The self-assembled structure consists of horizontally aligned rows which are connected via H-bonds. The on-surface chirality[36] of adenine can be distinguished in the enlarged high-resolution STM image (Figure 1b) with a CO-tip by their asymmetrically shaped structures. The corresponding constant-height AFM image (Figure 1c) further resolves the detailed structure of adenine molecules and their bonding network. The simulated AFM (Figure 1d), with a superimposed DFT relaxed model of the canonical form adenine, indicates that the adenine pair across the gap between rows is homochiral, though the whole network is racemic. The low-dimensional superstructure is stable at RT and it resembles the bulk crystal structure of adenine,[37,38] suggesting the intermolecular H-bonds are much stronger than the interaction with the substrate.

After exposure to water vapour at a pressure of $\sim 1 \times 10^{-5}$ mbar for 15 min at a temperature of $\sim 200$ K, some protrusions show up between rows of the superstructure in the high-resolution STM image (Figure 2a), originated from the insertion of water molecules.[39] Here, the right "hole" of the superstructure contains a water molecule and the simulated images correspond to this region. However, the constant-height AFM image looks similar to that without water (Figure 2b). The configuration of the adenine superstructure revealed by the high-resolution STM image is matched in the simulated result but the protrusions originating from the adsorbed water molecules are not seen in the simulation (Figure 2b). In the inset of Figure 2b, the simulated AFM image based on the most reasonable configuration of the water molecule (see Figure S2) matches well with the experimental result. Also, a similar orientation of a single water molecule in an adenine superstructure has been previously reported.[26] Overall, this implies that the configuration is likely correct, but that the model for STM simulations does not fully capture the distance dependence of the imaging in this case. The very faint contrast of the single water molecule in the AFM image is pre-sumably due to its position slightly lower than the adjacent adenine molecules. Overlaying the optimized structural model based on the DFT calculations (Figure 2b) reveals that the water molecule is stabilized by the surrounding adenine molecules through H-bond and the surface chirality of adenines remains unchanged. The distance between rows increases very slightly (see Figures S5, S6, and S8), suggesting that introducing water has a minor effect on the layer when the substrate is at ∼200 K.

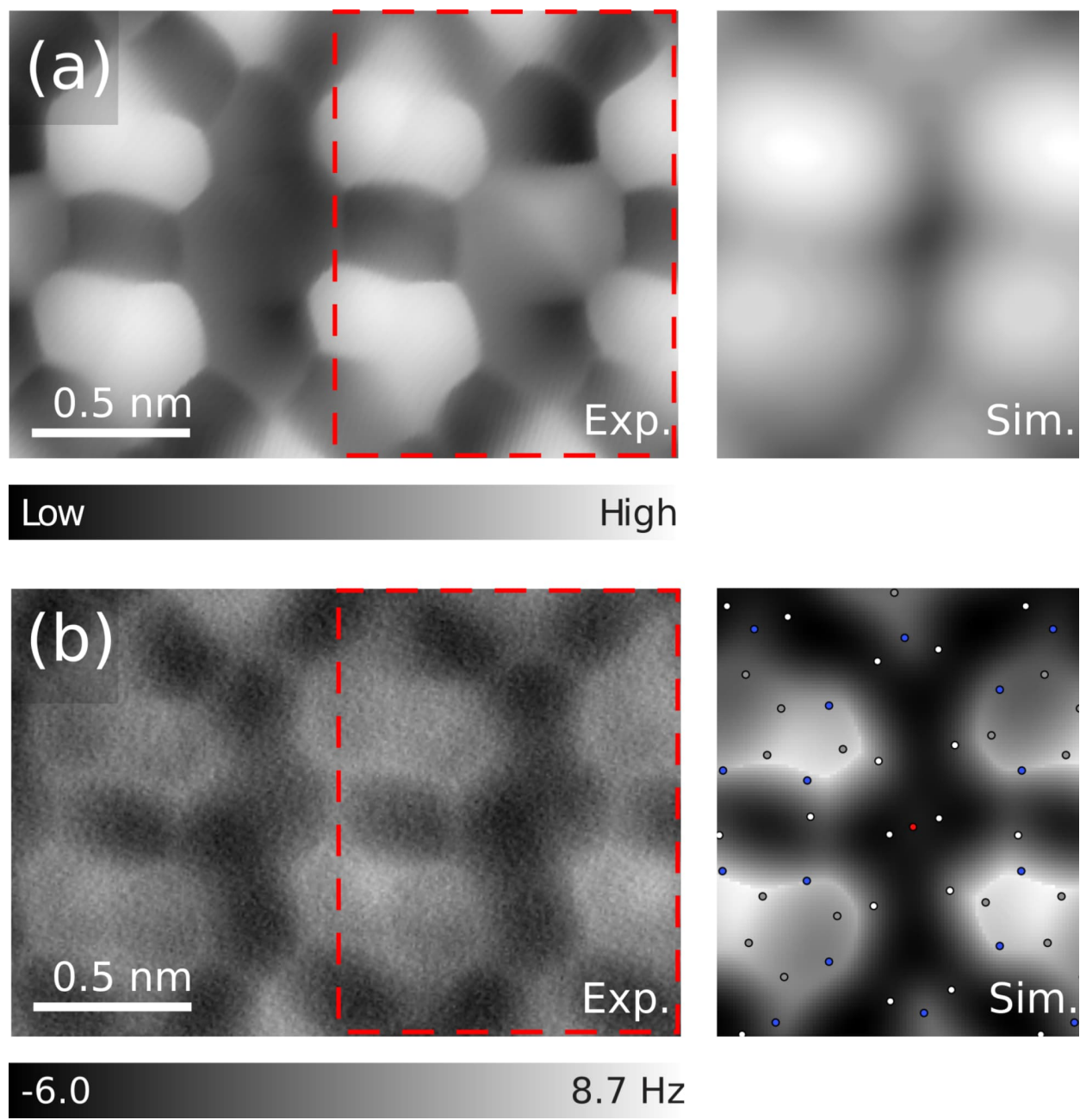

Figure 2: Adenine on Ag(111) after dosing with water molecules at low temperature. (a) STM topography showing water molecule(s) embedded among adenine molecules with a simulated STM image in the adjacent panel. Setpoint: 50 mV, 150 pA. (b) The corresponding constant-height AFM image with a simulated AFM image in the adjacent panel. DFT optimized adsorbate structure overlaid. Experimental images have been recorded with a CO functionalized tip apex.

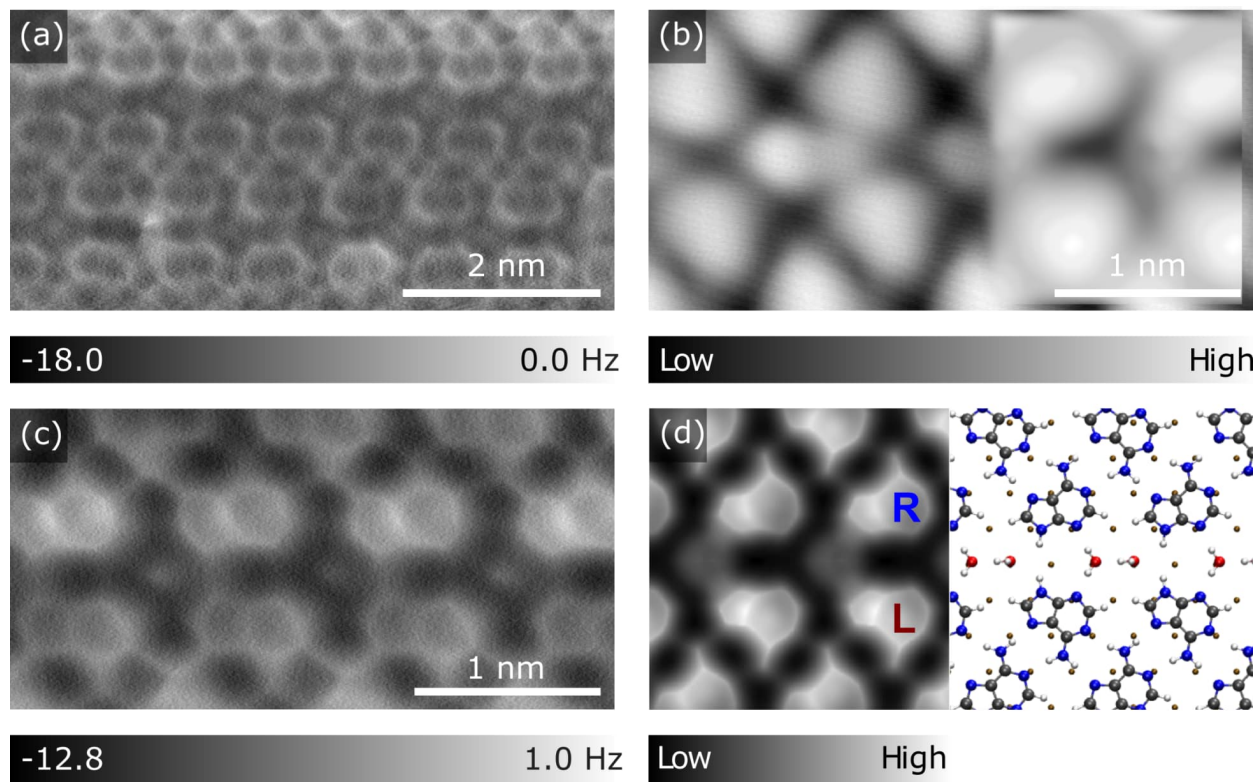

Figure 3: Annealed hydrated adenine structure on Ag(111) imaged with CO-tip. (a) Constant-current AFM image of an adenine island with water. (b) Zoomed-in STM topography of the island with a simulated STM image as an inset. Setpoint: 250 mV, 100 pA. (c) Constant-height AFM image of the zoomed-in area of the island. (d) Simulated AFM image of the hydrated adenine structure and corresponding molecular structure based on the DFT calculations.

After depositing water molecules on a sample that was held at RT, a water-involved rectangular superstructure, in contrast to the previous rhombic-shaped self-assembly, is obtained (see Figure S4). Also, some bright protrusion features and heterochiral adenine pairs across rows show up (Figure 3a). Noticeably, in the zoomed-in high-resolution STM image (Fig-ure 3b), the bright protrusion feature surrounded by two adenine pairs closely resembles the STM image of water dimers reported earlier.[40] The spatial arrangement of the heterochiral adenine pairs across rows indicates the conversion from matched into mismatched H-bonding pattern. The mismatched pattern is not found in the superstructure without water or with the water deposited when the substrate is at $\sim 200$ K. In the constant-height AFM image, a dot-shaped protrusion shows up in the corresponding area of the bright protrusion in the STM image and the distance between adenine rows is increased to about 0.82 nm (approximately 26% and 22% larger than that without water and with the water deposited when the substrate is at $\sim 200$ K, respectively, see Figures S7 and S8). The simulated AFM image based on the proposed water dimer model with one flat and one upright water molecule agrees well with the experimental result as shown in Figure 3c. As depicted in the optimized structure model (right part in Figure 3d), two N-H moieties in the adenine pairs point to each other and form an unusual mismatched H-bonding pattern between the molecules instead of the ordinary N-H$\cdots$N bond.

Figure 4 compares the calculated adenine self-assembly structures with the usual H-bonding pattern as well as the mismatched H-bonding configuration with and without the presence of water dimers. The structure with linear non-planar (LNP) water dimers has the lowest energy. Other water dimer structures, *e.g.* a linear planar form where all four hydrogen atoms and two oxygen atoms lie in the same plane have a higher total energy. Similarly to the LNP water dimer in our self-assembled structure, the linear non-planar water dimer is the minimum energy structure also in the gas phase.[41] For the water dimer in the adenine layer, the O–H-O bond angle is 171.5° and the H-bond is 1.81 Å in length (compared to 175.5° and 1.93 Å for an LNP dimer in the gas phase). The estimated energy gain from bonding to the water dimer $E_b$ before (NH$\cdots$N) and after hydration (NH$\cdots$N and OH$\cdots$O) is $-2.6$ eV and $-3.5$ eV per unit cell, respectively, indicating that loss from the mismatched H-bonding is more than offset by the interaction with the water dimer (see SI). The energy difference between the mismatched hydrogen bond network with and without water dimers is approximately 1.2 eV per unit cell, clearly showing that a water dimer is necessary for stabilizing this unique structure. Finally, our experimental results and DFT calculations point to the presence of water dimers and not monomers in the self-assembled adenine layer. This observation is further supported by the fact that water dimers are predicted to be a better hydrogen donor and acceptor compared to water monomers,[42] which also implies stronger interactions between the dynamically self-assembled layer and the spontaneously formed water dimers.

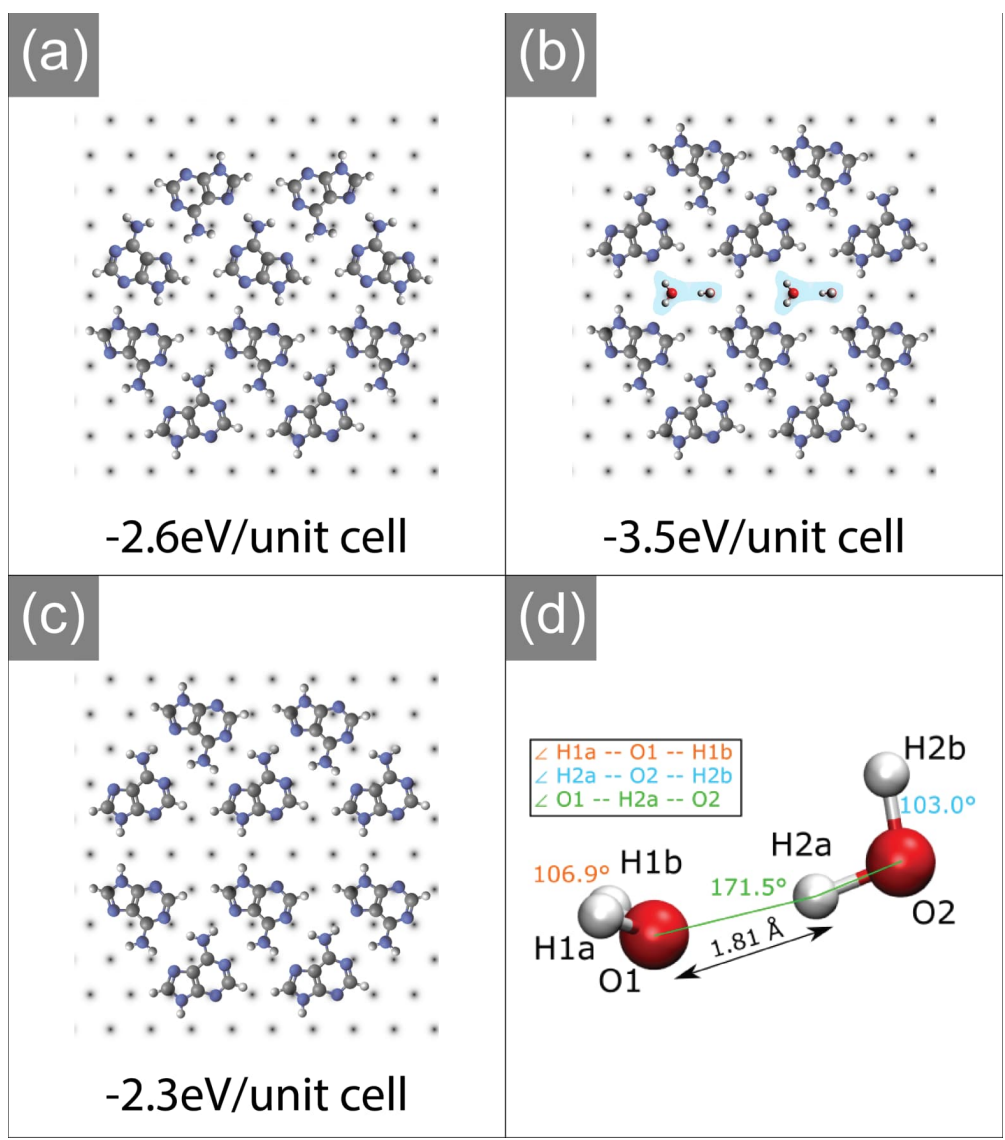

Figure 4: (a-c) The structural models of the self-assembled adenine structures without (a,c) and with water dimers (b). The panel (a) shows the adenine layer with the H-bond matched structure, while panels (b,c) depict the H-bond mismatched structure. (d) Detailed structure of the linear non-planar water dimer found in the adenine layer.

To rule out the possible tautomeric behaviour of adenine in the self-assembled layer, we performed further DFT calculations and AFM simulations on possible structures as previous research indicates that other tautomers besides the canonical form might appear at RT in a microhydrated environment.[43] However, we found no concrete evidence suggesting that secondary tautomers are found in the hydrated supramolecular network, and the canonical form adenine molecule should be the primary component.

In conclusion, by introducing water to the self-assembled adenine layers at RT, individual water dimers were formed and stabilized inside the restructured adenine 2D network. With STM, AFM and DFT simulations, we successfully revealed the detailed bonding structure of the confined water dimers as well as the re-arrangement the local adenine network undergoes upon hydration. The water dimers under confinement appear to be in a linear non-planar configuration, causing a local chirality inversion such that a distinctive mismatched H-bond pattern emerged between neighboring adenine molecules. The comprehensive characterization of the ensemble of water dimers and adenine not only provides crucial insights into the dynamic nature of the hydration process of DNA bases but also offers a novel method to study unstable small molecule clusters which would otherwise be impossible to observe.

# Acknowledgement

This research made use of the Aalto Nanomicroscopy Center (Aalto NMC) facilities and was supported by the European Research Council (ERC 2017 AdG no. 788185 "Artificial Designer Materials"), and the Academy of Finland (projects no. 347611, 347319, 346824, 314862, Centres of Excellence Program project no. 284621, and Academy professor funding no. 318995 and 345066). A.S.F. has been supported by the World Premier International Research Center Initiative (WPI), MEXT, Japan. C.X. acknowledges funding from the European Union's Horizon 2020 research and innovation programme under the Marie

Skłodowska-Curie grant agreement "EIM" No 897828. This work was undertaken as part of the FinnCERES competence centre. Computing resources from the Aalto Science-IT project and CSC, Helsinki are gratefully acknowledged.

## Supporting Information Available

Further experimental and computational details and results.

# Supporting Information: Water dimer driven DNA base superstructure with mismatched hydrogen-bonding

## Methods

**Experimental**

The experiments were carried out on a combined non-contact AFM/STM system (CreaTec) with a commercial qPlus sensor with a Pt/Ir tip, operating at $T \approx 5$ K in ultrahigh vacuum at a pressure of $\sim 1 \times 10^{-10}$ mbar. The qPlus sensor had a resonance frequency of $f_0 \approx 30046$ Hz, a quality factor $Q \approx 67714$, and was always operating with an oscillation amplitude of $A = 50$ pm.

The Ag(111) substrate (MaTeck) was prepared by repeated Ne$^+$ sputtering with a beam energy of 1000 eV and ion current of 30 µA for 15 min followed by annealing at $\sim 480 - 500$°C for 5 min. A flat Ag(111) surface with large terrace and minimum amount of impurities was obtained within 3 cycles. The adenine molecules (Sigma-Aldrich; purity $\geq 99\%$) were deposited onto the substrate at $\sim 5$ K through thermal sublimation at 120°C at a chamber pressure of $1 \times 10^{-6}$ mbar for 15 min. A self-assembly layer of adenine can be formed by slightly warming up of the substrate for 5 min. Then the CO molecules (Praxair; purity 99.997%) were deposited onto the substrate at $\sim 5$ K through a variable leak valve at a chamber pressure of $1 \times 10^{-8}$ mbar for 80 seconds.

Before deposition, the adenine was thoroughly degassed at 120°C for 1 h. While the water was firstly boiled at 100°C to rid of any residual gas inside, and was then degassed thoroughly via several freeze-pump-thaw cycles. [1]

During the hydration process, the sample was held at $T \approx 200$ K or room temperature, and then was transferred to the preparation chamber. The water molecule (Sigma-Aldrich SKU38796; deionized) was introduced into the preparation chamber through a variable leak valve aiming directly at the sample. The pressure was maintained at $\sim 1 \times 10^{-5}$ mbar for 15 min. The sample was later transferred back to the main chamber and cooled down to 5 K for imaging.

## Computational

**DFT calculations.** All calculations were performed using the all-electron density functional theory code FHI-AIMS [2, 3] with the PBE exchange correlation functional [4] and the Tkatchenko-Scheffler method [5] was used to accurately handle the van der Waals interactions. The basis sets as defined in FHI-AIMS were used at the "light" level for all atoms. A trust radius enhanced version of the BFGS algorithm was used to relax the structures to a force less than $10^{-2}$ eV/Å. To acquire an approximately homogeneous k-grid in the plane parallel to the substrate, a $7 \times 2 \times 1$-grid was used.

As Fig. S1 shows, the adenine monolayer was modeled as a pair of adenine chains that are periodic only in the horizontal direction. Still, one gap across the chains is retained and the simulated AFM images in this region show good correspondence to the experimentally acquired images. The choice to exclude the periodicity in the vertical direction was made as there was a significant lattice mismatch between the Ag(111) substrate and the adsorbed adenine monolayer. In the horizontal direction, the lattice mismatch was less than 2 % in all structures. As the inclusion of water mostly affected the vertical gap between the chains and not the horizontal, we could use the same unit cell dimensions in all calculations. Three substrate layers were used in all calculations. The DFT relaxed hydrated layers modelling the experiments at $T = 200\,\text{K}$ and at room temperature are shown in Fig. S2 and Fig. S3 respectively.

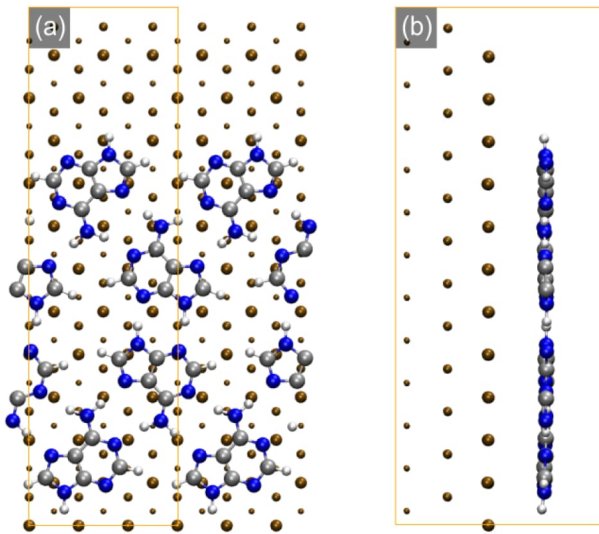

Figure S1: (a) The DFT relaxed structure of the pure A layer in top view with one unit cell ($8.63\,\text{Å} \times 29.9\,\text{Å} \times 30.0\,\text{Å}$) highlighted. The three layers of the substrate are illustrated with spheres of different radii. (b) The relaxed structure in side view.

**AFM simulations.** The probe particle model [6, 7] was used to produce the simulated AFM images. To emulate the behaviour of the CO tip, we used a lateral spring constant of $0.24$ N/m and a radial spring constant of $20.00$ N/m for the probe particle. Further, the tip charge was modeled as a quadrupole with a quadrupole moment of $-0.05\,\text{e} \times \text{Å}^2$. These are standard parameters for CO tip imaging [6, 7]. The electrostatic field of the sample was obtained using FHI-AIMS.

**STM simulations.** All simulated STM images were obtained using the PP-STM code [8]. The scans were performed using the relaxed tip positions acquired during the AFM simulation. The broadening factor was set at $0.5$ eV and standard CO tip orbital parameters were used with $13\%$ s orbital and $87\%$ $p_{xy}$ orbital contributions [9]. The constant-current images were produced by calculating an isosurface from a stack of constant-height images at an isovalue of approximately 10% of the maximum current. For all structures, the constant-height stack was calculated between 5 Å and 7 Å above the sample. A 0.1 Å spacing was used for the constant-height scans.

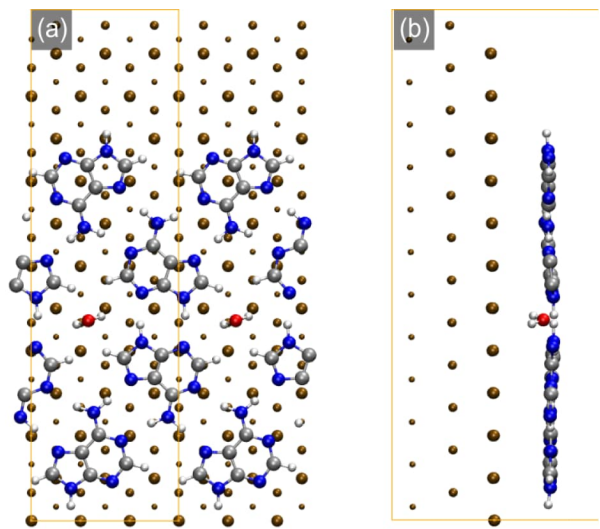

Figure S2: (a) The DFT relaxed structure of the hydrated A layer modelling experiments at $T \approx 200\,\text{K}$ in top view with one unit cell highlighted. (b) The relaxed structure in side view.

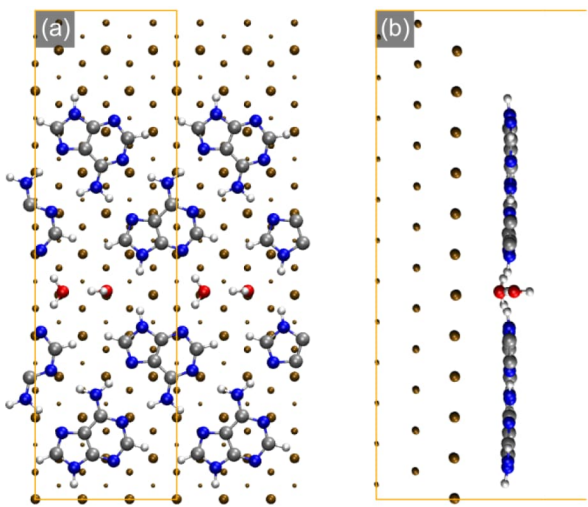

Figure S3: (a) The DFT relaxed structure of the hydrated A layer modelling experiments at room temperature in top view with one unit cell highlighted. (b) The relaxed structure in side view.

# Overview STM images

Fig. S4 shows large scale overview images of the adenine self-assembly structure for non-hydrated and hydrated and annealed samples.

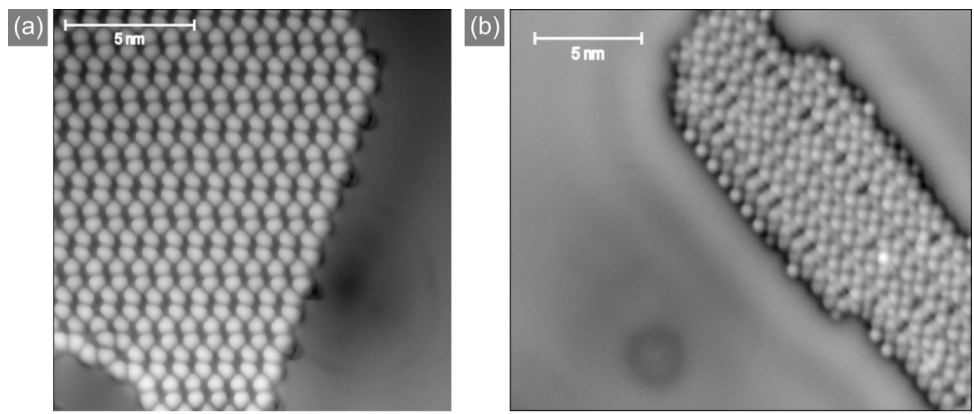

Figure S4: STM images with CO-tip showing large overview of (a) a non-hydrated assembly and (b) a hydrated adenine assembly after annealing at RT.

# Determination of the lattice parameters of the adenine structures

We have used line profile analysis of the STM images to determine the spacing between the adenine rows in the three different samples (non-hydrated, hydrated and annealed to 200 K, and hydrated and annealed to RT), see Figs. S5-S7. The analysis can be complemented by using 2D autocorrelation as shown in Fig. S8, which allows determining the lattice parameters of the structures. This can be converted to the distance between the pairs of adenines as indicated in the figure.

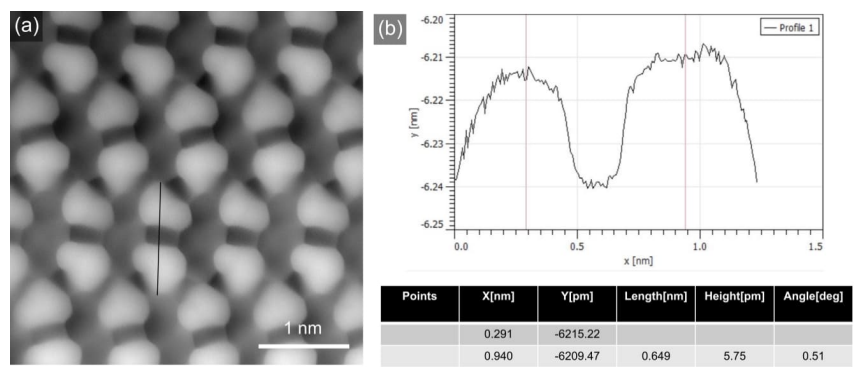

Figure S5: Example of the distance determination between the adenine rows before hydration. Average over several line profiles yields 0.65 nm.

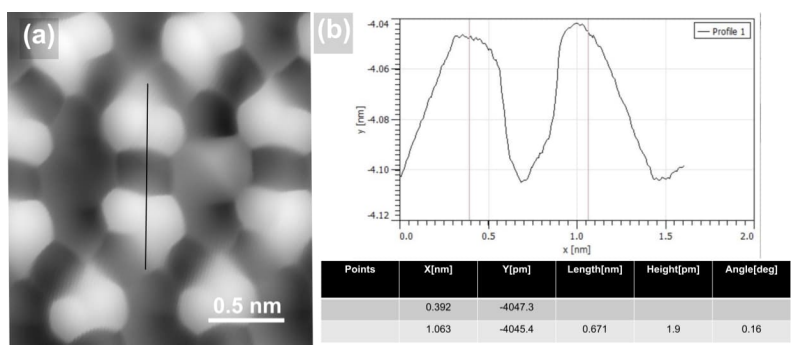

Figure S6: Example of the distance determination between the adenine rows after hydration at 200 K. Average over several line profiles yields 0.67 nm.

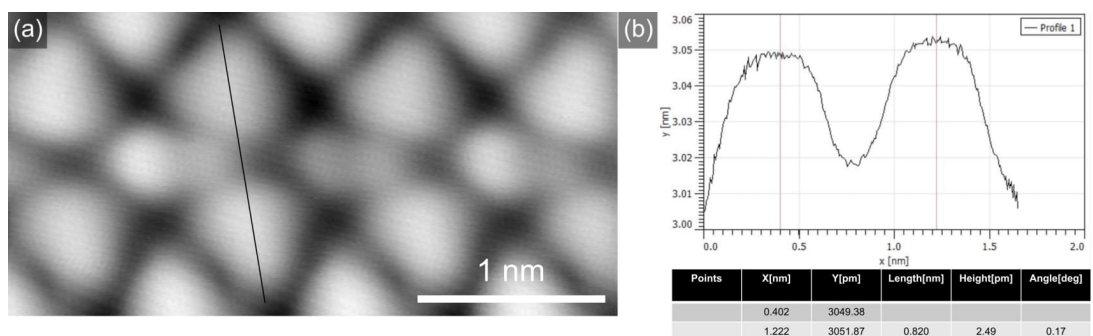

Figure S7: Example of the distance determination between the adenine rows after hydration at RT. Average over several line profiles yields 0.82 nm.

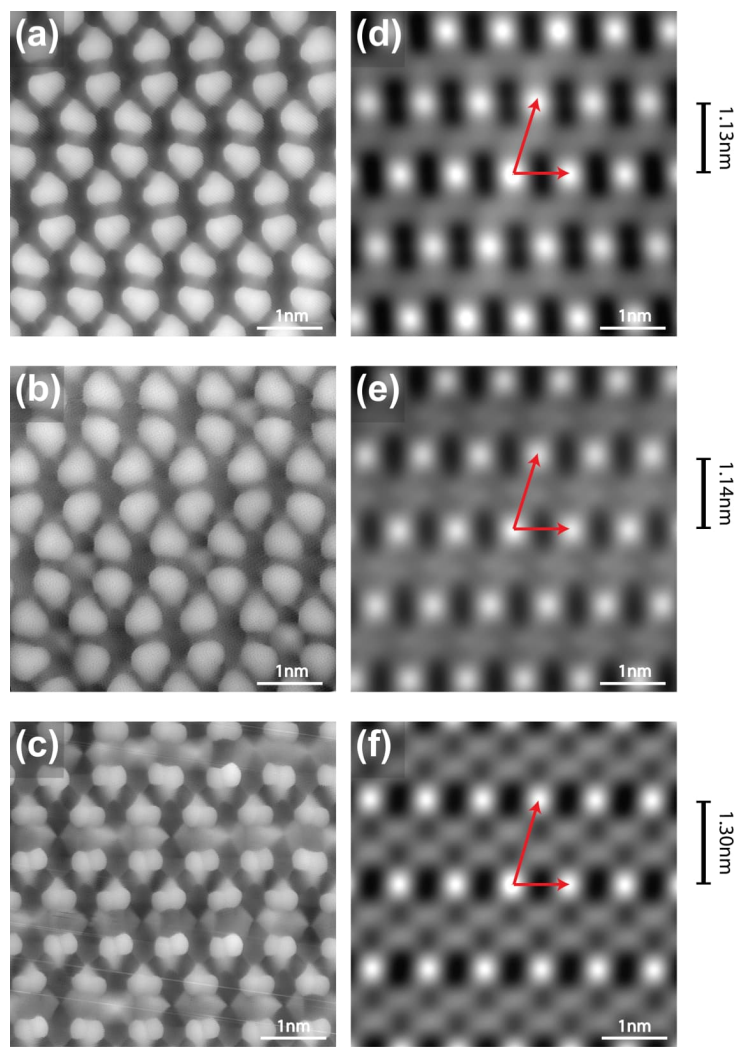

Figure S8: Measuring the lattice parameter of the self-assembled adenine structures. (a), (b) and (c) STM topography with CO-tip showing adenine layer without water, after hydration at 200K and after hydration at RT, respectively. (d) 2D auto-correlation[10] of (a), the distance between un-hydrated adenine rows is 1.13 nm, which is calculated from the lattice of the 2D auto-correlation, marked by the red arrows. Similarly for (e) and (f), the distance for adenine rows after hydration at 200K is 1.14 nm and for after hydration at RT, the distance becomes 1.30 nm.